\documentclass[aip,graphicx,reprint]{revtex4-1}

\usepackage{graphicx}
\usepackage{dcolumn}
\usepackage{bm}

\usepackage{comment}
\usepackage{amssymb}
\newcommand{\LoopGain}{\mathcal{L}}
\usepackage{placeins}
\let\Oldsection\section
\renewcommand{\section}{\FloatBarrier\Oldsection}
\let\Oldsubsection\subsection
\renewcommand{\subsection}{\FloatBarrier\Oldsubsection}
\let\Oldsubsubsection\subsubsection
\renewcommand{\subsubsection}{\FloatBarrier\Oldsubsubsection}

\draft 

\begin{document}

\title{Characterization and Improvement of the Thermal Stability of TES Bolometers} 

\author{Rita Sonka}
\affiliation{Department of Physics, California Institute of Technology, Pasadena, California 91125, USA}
\author{James J. Bock}
\affiliation{Department of Physics, California Institute of Technology, Pasadena, California 91125, USA}
\affiliation{Jet Propulsion Laboratory, Pasadena, California 91109, USA}
\author{Krikor G. Megerian}
\affiliation{Jet Propulsion Laboratory, Pasadena, California 91109, USA}
\author{Bryan A. Steinbach}
\affiliation{Department of Physics, California Institute of Technology, Pasadena, California 91125, USA}
\author{Anthony D. Turner}
\affiliation{Jet Propulsion Laboratory, Pasadena, California 91109, USA}
\author{Cheng Zhang}
\affiliation{Department of Physics, California Institute of Technology, Pasadena, California 91125, USA}

\date{\today}

\begin{abstract}
We study the mechanism of instability in transition edge sensor (TES) bolometers used for ground based observations of the Cosmic Microwave Background (CMB) at 270GHz.
The instability limits the range of useful operating resistances of the TES down to $\approx$ 50\% of $R_n$, and due to variations in detector properties and optical loading within a column of multiplexed detectors, limits the effective on sky yield.
Using measurements of the electrical impedance of the detectors, we show the instability is due to the increased bolometer leg $G$ for higher-frequency detection inducing decoupling of the palladium-gold heat capacity from the thermistor.
We demonstrate experimentally that the limiting thermal resistance is due to the small cross sectional area of the silicon nitride bolometer island, and so is easily fixed by layering palladium-gold over an oxide protected TES.
The resulting detectors can be biased down to a resistance $\approx$10\% of $R_n$.

\end{abstract}

\maketitle 

\section{Introduction}

\begin{figure}
\includegraphics[width=3.37in]{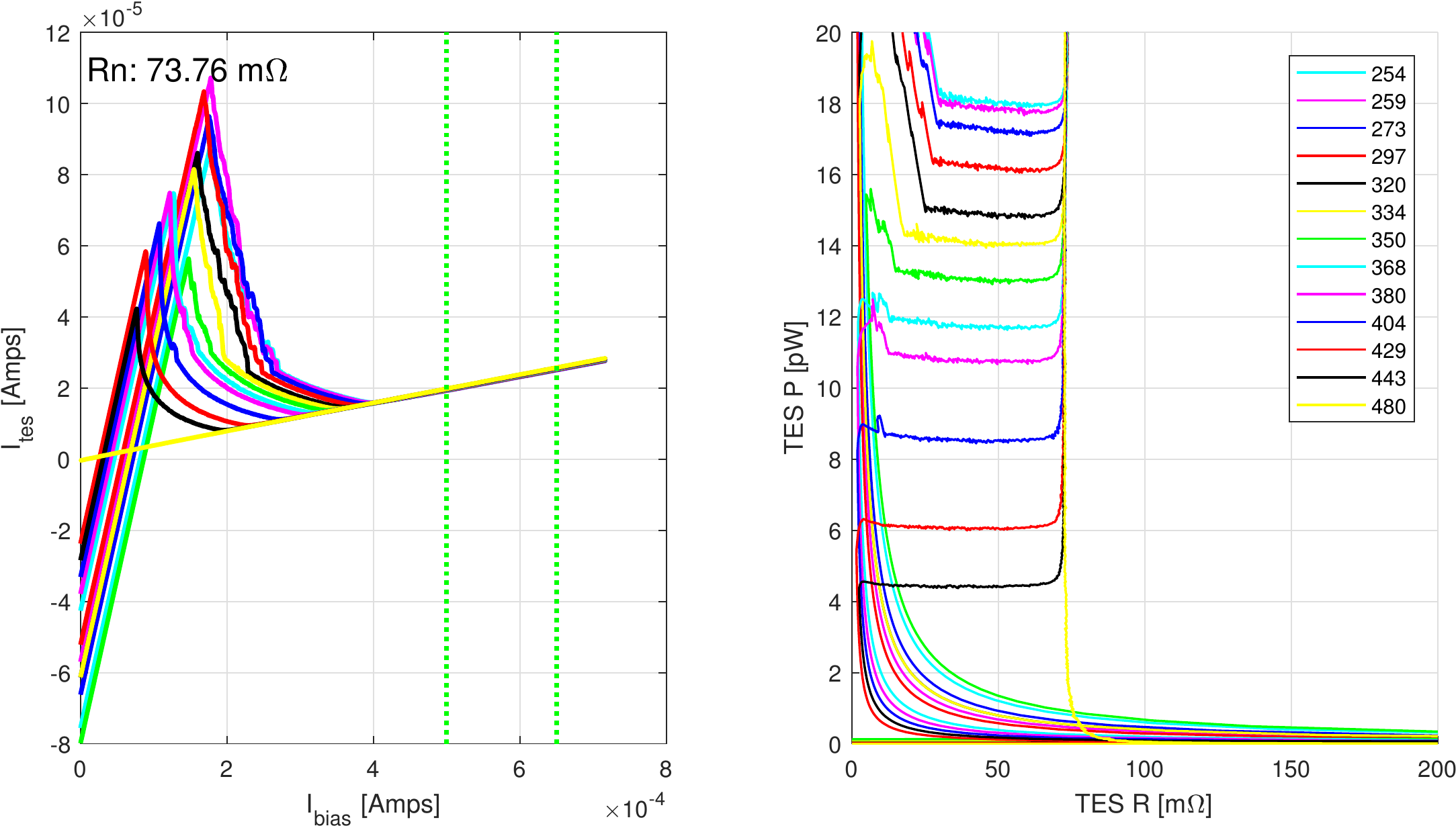} 
 \caption{\label{fig:iv_col_0_row_9_type_A_ex} Original (type A) detector characteristic IV and PR plots. } 
\end{figure}

The successful detection and characterization of the B-modes in the Cosmic Microwave Background (CMB) would dramatically illuminate the physics of the inflationary era.
CMB measurements are currently limited by a foreground of galactic dust\cite{planck_collaboration_planck_2016}\cite{keck2016} , which can be subtracted by imaging at higher frequencies where the CMB is dim such as the 200-300GHz atmospheric band.
The state of the art detector astrophysical imaging in the millimeter waves is the Transition-Edge Sensor (TES) bolometer\cite{irwin_hilton_2005}.
The bolometer thermal conductivity $G$ used for 270GHz imaging in a ground based experiment must be higher than that used for 95,150 or 220GHz due to the increased atmosphere temperature.
In detectors designed for the Keck experiment at 270GHz, we found that the stable operating region for these higher $G$ detectors was limited, and the detectors would latch if operated at fractional resistances less than $\approx 50\%$ of the normal resistance, as shown in Fig. ~\ref{fig:iv_col_0_row_9_type_A_ex}.

While individually our detectors have high yield, when they are multiplexed into a large format array using a time multiplexing system, the yield is limited by the requirement that a column of 32 detectors is DC biased with a single bias voltage.
Variations in detector properties and optical loading across the focal plane cause variations in the optimal bias point for the detectors.
In the lower frequency instruments, the stable regions were wide enough that the constant voltage within a column requirement did not impact yield.
In contrast, the limited stability of the 270GHz detectors precluded operation of all the detectors in a column at a single bias.
In the dark measurements presented here, even with no optical power variations, the simultaneously biased into transition yield was at best 67\% for the baseline detector design, as shown in Fig. ~\ref{fig:trg_type_A}.

\begin{figure}
\includegraphics[width=3.37in]{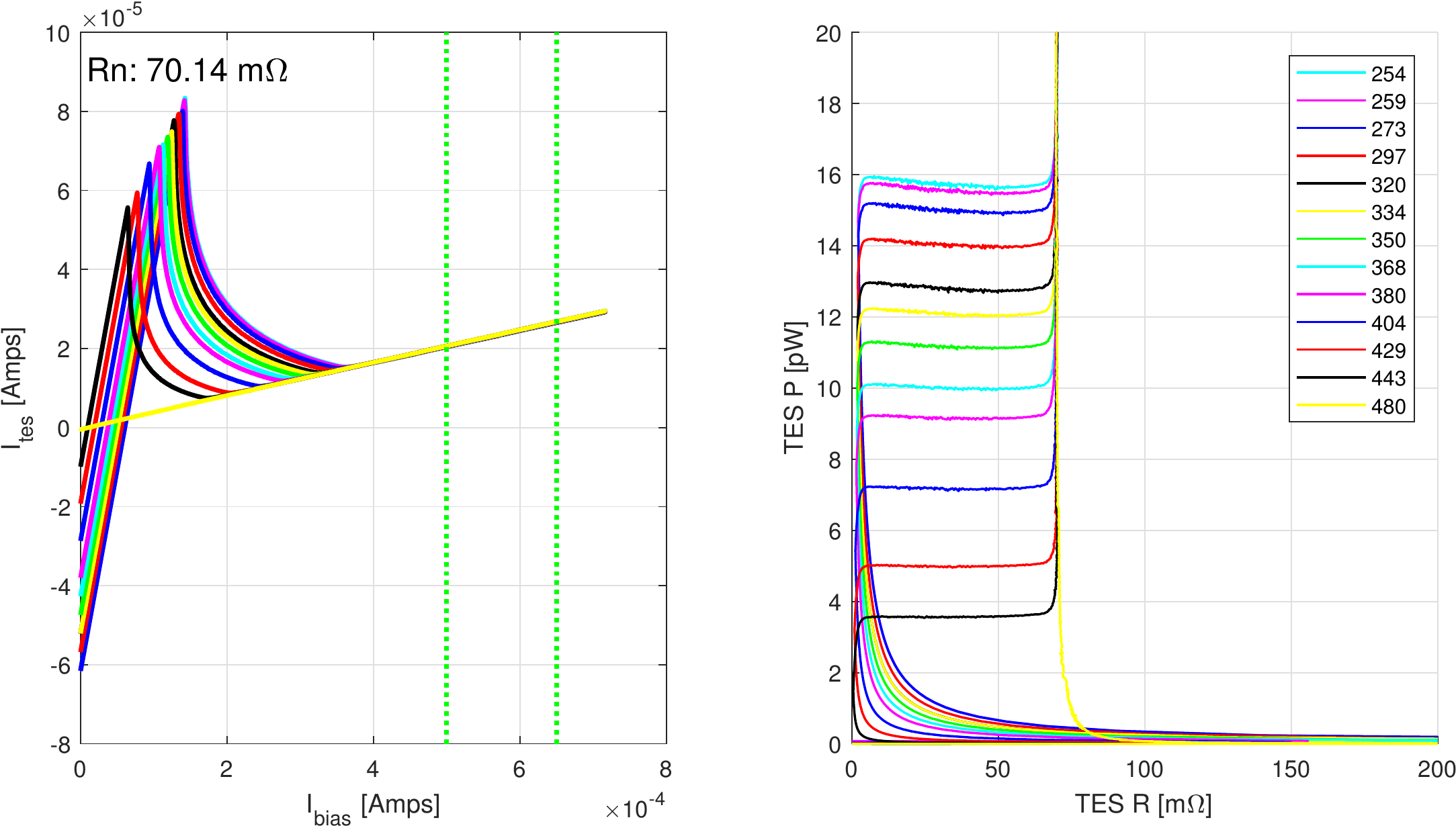} 
 \caption{\label{fig:iv_col_0_row_19_type_F_ex} Fully functional (type F) detector IV and PR plot.  } 
\end{figure}

\begin{figure} 
\includegraphics[width=3.37in]{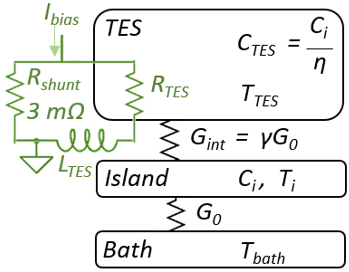} 
\caption{\label{fig:BolometerTheoreticalModel} Bolometer thermal (black) and electrical (green) circuit.}
\end{figure}

Inspired by the work of George et al \cite{EGeorge2014}, we suspected that the stability of the detectors was limited by the TES-heat capacity coupling, rather than limited electrical bandwidth\cite{irwin_hilton_2005}.
In this model, detector stability requires that the ratio $\gamma$ of thermal conductivity within the island between the titanium TES and the palladium-gold (PdAu) heat capacity to G be greater than the electrothermal feedback loopgain.
Increasing the detector G for higher frequency detectors without changing the island design decreased $\gamma$ to a point where it fell below the loopgain achievable from the steepness of the superconducting transition.
The thermal circuit of this model is depicted in Fig.~\ref{fig:BolometerTheoreticalModel}, and leads to the following stability criterion:

\begin{eqnarray}
\LoopGain < \gamma + 1 + \frac{C_{TES}}{G \frac{\gamma}{\gamma + 1}\tau_{e}} \approx \gamma
\label{eq:3rdStability}.
\end{eqnarray}

$C_{TES}$ is the heat capacity of the titanium TES and $\tau_e$ is the electrical time constant of the readout circuit $L/R$.
The term $\frac{C_{TES}}{G \frac{\gamma}{\gamma + 1} \tau_e}$ is negligible because $C_{TES} << C_i$, so $\frac{C_{TES}}{G} << \frac{C_i}{G} = \tau_0 < \tau_e$, where $C_i$ is the heat capacity of the bolometer island.

\section{Bolometer decoupling model}

We considered two possible mechanisms for the internal thermal resistance of the island: Thermal resistance due to the small cross sectional area of the island the TES and PdAu are mounted on, and a thermal resistance between the quasiparticles in the superconductor and the phonons.

The thermal conductance of a rectangular aperture of an insulator reaches a limiting value as the insulator gets thinner, determined by the Stefan-Boltzmann law for radiation.
The limiting value in silicon nitride has been shown to be consistent with
this model\cite{holmes1998}\cite{woodcraft2000}:

\begin{equation} G = 4 \Sigma A T^3
\label{eqn:limitingThermalConductance}
\end{equation}

Where G is the thermal conductance $dP/dT$, $\Sigma$ is the Stefan-Boltzmann constant for phonons in silicon nitride of 15.7$mW/cm^2 K^4$, and $A$ is the cross sectional area.
The cross sectional area of our island is $\approx$150 $\mu m^2$, and all the heat between the TES and the PdAu heat capacity must pass through this small aperture for the baseline A type detector.
For the titanium transition critical temperature $T_c=0.5K$, this is a decoupling thermal conductance of $G_{int} = 12 nW/K$.
For a typical leg thermal conductivity $G \approx 150pW/K$, $\gamma \approx 80$, and stable detector operation will be limited to loop gain $\mathcal L < 80$.

The heat capacity of the titanium TES we estimate at $C_{TES} = 0.016pJ/K$ from its volume, 100$\mu m^3$, the critical temperature of our titanium films $T_c=0.50K$, and the bulk electronic heat capacity of titanium, 310 $J m^{-3} K^{-2}$.
The island heat capacity we estimate $C_i = 4.8pJ/K$, from the measured bolometer $\tau_0$ and $G$.
Because $C_{TES}$ is so small compared to the $C_i$, we neglect including it in the model for the bolometer impedance.

A second possibility for the thermal resistance is the electron-phonon coupling, i.e. hot electron effects.
Measurements of the electron phonon relaxation time in titanium films\cite{gershenson2001}\cite{wei2008} give 1-3 microseconds at 0.5K, though in films with $T_c < 0.5K$.
Given a TES volume of $\approx 100 \mu m^3$, $G_{e-ph} = 6-17nW/K$, or $\gamma = 40-110$.
$G_{e-ph}$ and $G_{int}$ are quite similar but we can distinguish between them by modifying the bolometer to bring the palladium-gold closer to the TES with or with out direct electrical contact.

\section{Bolometer modifications}

\begin{figure}
\includegraphics[width=3.37in]{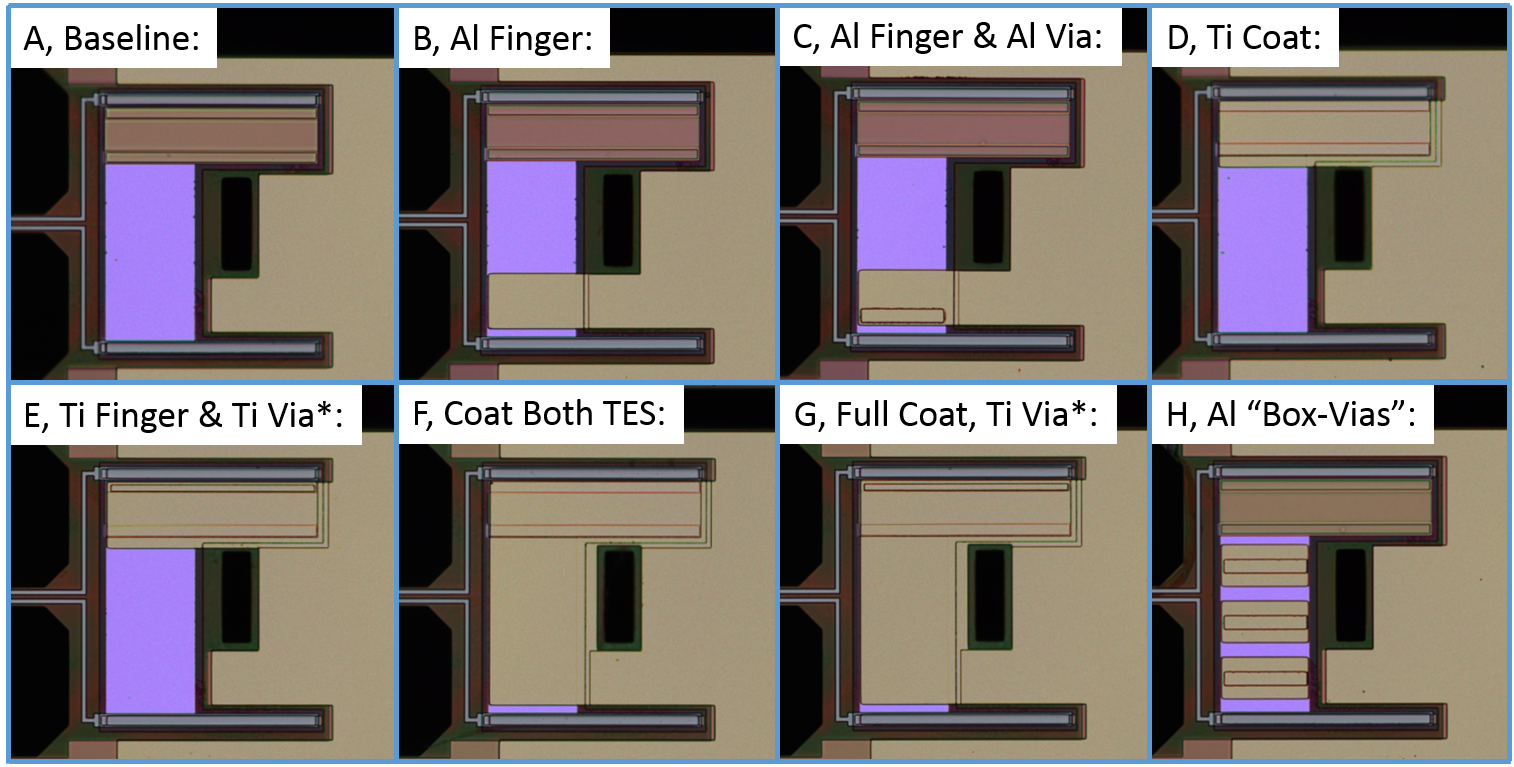} 
\caption{\label{fig:designDiagrams} Photographs of the 8 bolometer designs.}
\end{figure}

To test if internal thermal conductivity was responsible for limiting our detector stability, we manufactured a wafer of detectors with eight modifications to the layout of the TES and heat capacity.
Fig. ~\ref{fig:designDiagrams} shows the eight different styles we tested.
They explore different connections to the aluminum TES and titanium TES.
In this paper we investigate the effect of the modifications only on the titanium TES, which is the sensor used for science observations.
\begin{itemize}
	\item A - The baseline detector.
	\item B - A finger of palladium-gold is extended over the aluminum TES ($T_c=1.2K$), but not the titanium TES, separated by oxide.
	\item C - Like B, but with a via through the oxide to allow direct electrical contact between palladium-gold and aluminum.
	\item D - A finger of palladium gold is extended over the titanium TES, separated by oxide.
	\item E - Like D, but with a via through the oxide down to the titanium.
	\item F - Palladium-gold covers both aluminum and titanium TES, separated by oxide.
	\item G - Like F, with a via through the oxide down to the titanium.
	\item H - Short sections of palladium-gold with vias down to the aluminum.
\end{itemize}

One wafer of 128 detectors was fabricated with sixteen bolometers of each of these types.

\section{Results and Discussion}

\subsection{IV curves}

We first show the stability of the detectors through current-voltage (IV) curves and power-resistance (PR) curves.
The detector bias is set very high, to bring the detector normal, then the voltage is stepped down to zero.
An ideal detector has three regions in its IV and PR curves, corresponding to normal at high bias, in transition at intermediate bias, and superconducting at low bias.
Detector instability appears as an additional region between in transition and superconducting, where the readout system fails to track the detector current.
We estimate the range over which the detector is in transition by placing a criterion on the flatness of the PR curve, and looking at the difference between the normal resistance and the highest resistance where the detector is unstable.
We additionally cut from analysis detectors that have normal resistances far from the median, or large residuals to a straight line in the normal region.

\subsection{Transition range graphs}
\begin{figure}
\includegraphics[width=3.37in]{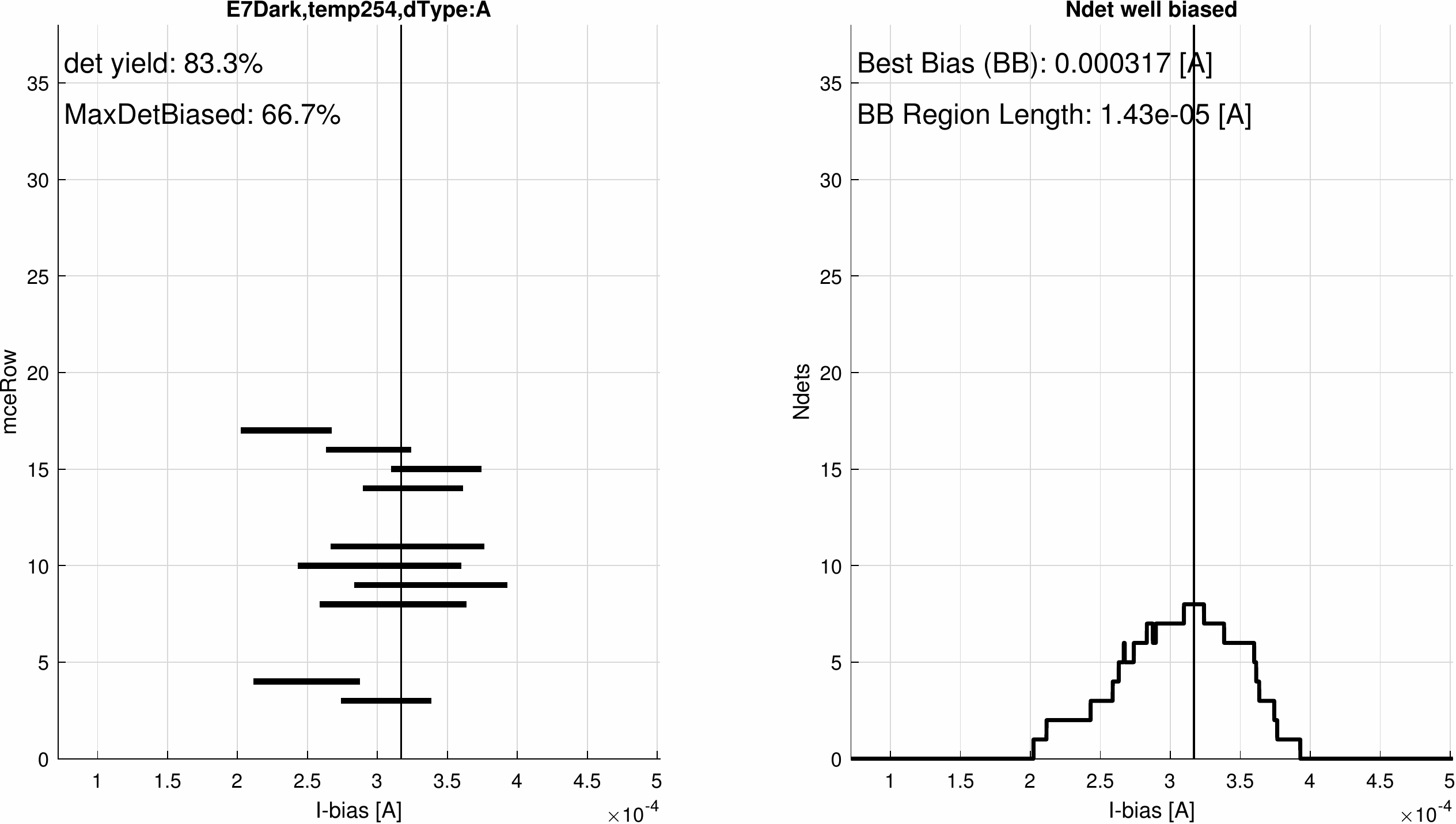} 
	\caption{\label{fig:trg_type_A} Type A transition range graph at a bath temperature of 254mK.  The x axes are the current applied to the TES shunt resistors.  The vertical line shows the optimal current where the largest number of detectors are operating simultaneously.  In the left plot, horizontal bars show the  limits of the in transition for each detector of type A.  The right plot shows the count of in transition detectors for each bias current.}
\end{figure}

\begin{figure}
\includegraphics[width=3.37in]{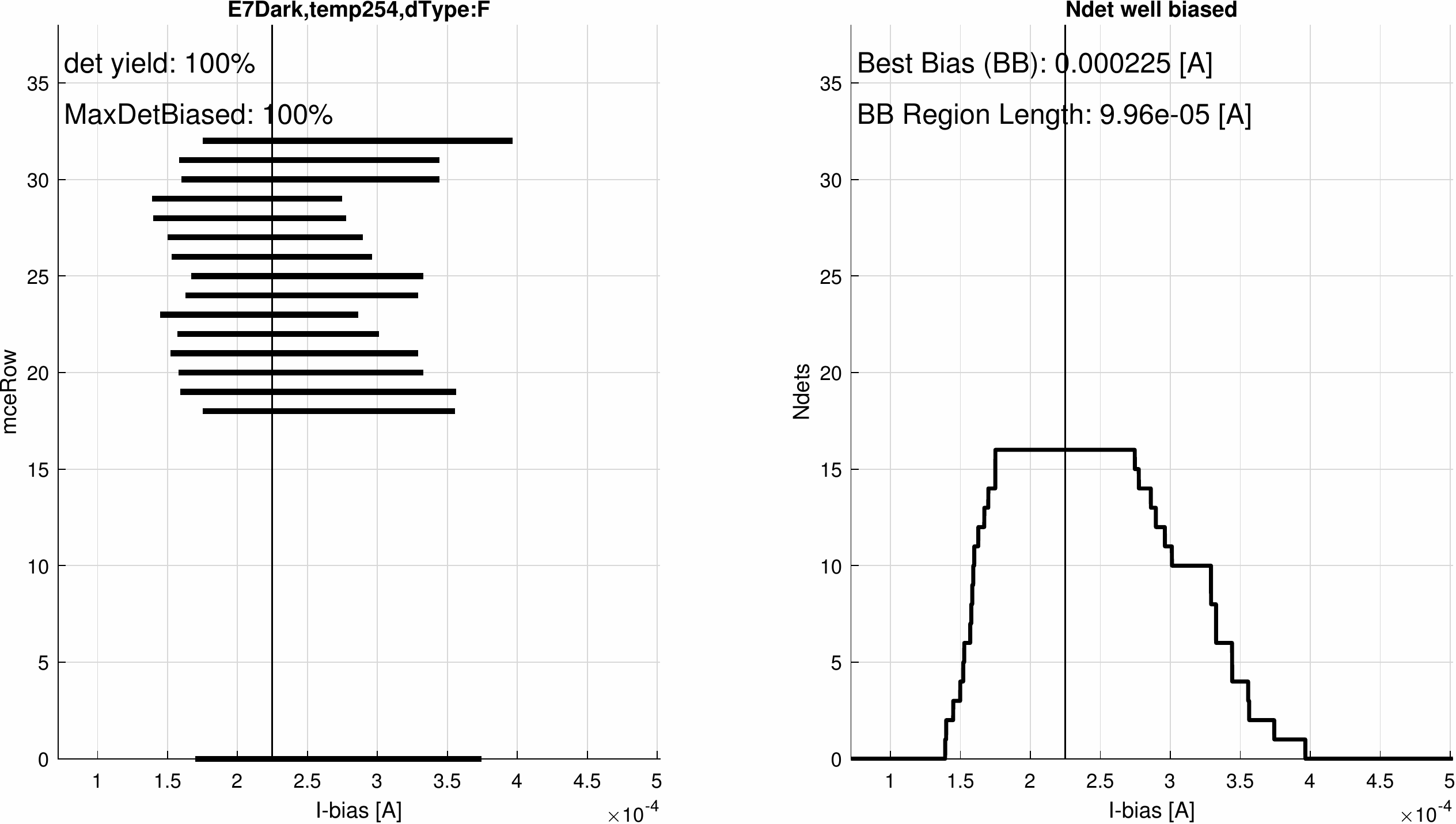} 
	\caption{\label{fig:trg_type_F} Type F transition range graph at a bath temperature of 254mK.  Compared to Fig.~\ref{fig:trg_type_A}, the stable bias regions are much wider for each detector.} 
\end{figure}

Table ~\ref{tab:detStats} displays the critical metrics for simultaneous operation of columns of detectors derived from the data for all detector types.
Figures ~\ref{fig:trg_type_A} and ~\ref{fig:trg_type_F} plot the transition ranges of detector types A and F and illustrate the derivation of the critical metrics.
The voltage bias in the table has units of microamps of current applied to the
3$m\Omega$ shunt resistor.

\begin{table}
\centering
\caption{Detector Stats}
\label{tab:detStats} 
\begin{tabular}{cccccc}
Det. & Det.  & Max Det. & Best 
&   Best Bias     & Mean  \\

type & yield\footnote{\% of detectors that displayed functional IV curves, not considering detectors where the readout system failed for reasons unrelated to the detectors.} & Simultaneously   & Bias 
&    Region Length & Length \\
    &     &   Biased\footnote{\% of simultaneously-biasable detectors that displayed functional IV curves, not considering detectors where the readout system failed for reasons unrelated to the detectors.} &   [$\mu A$] & [$\mu A$]  & [$\mu A$] \\
\hline
A     & 83     & 67     & 317     & 14     & 105     \\
B     & 88     & 69     & 280     & 4     & 100     \\
C     & 92     & 83     & 280     & 29     & 120     \\
D     & 93     & 93     & 247     & 93     & 170     \\
E     & 46     & 46     & 246     & 128     & 190     \\
F     & 100     & 100     & 225     & 100     & 169     \\
G     & 31     & 31     & 272     & 166     & 178     \\
H     & 100     & 100     & 313     & 25     & 126     \\
\end{tabular}
\end{table}

\begin{figure}
\includegraphics[width=3.37in]{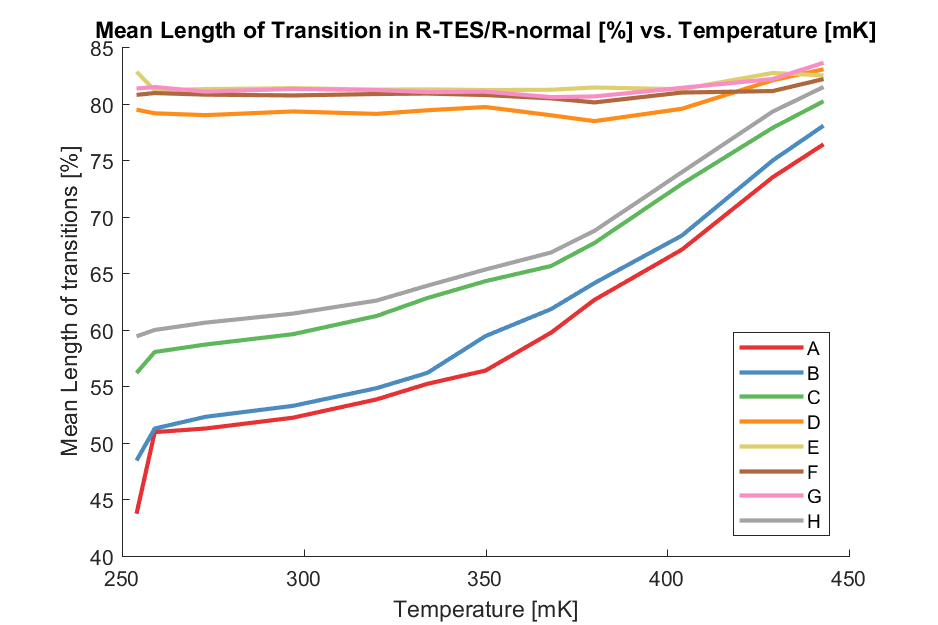} 
	\caption{\label{fig:meanLengthVsTemperature} Mean length in which a detector was usable (in $R_{TES}/R_n$ \%) vs. Temperature (in mK). Higher is better. Types A, B, C and H increase in stability with temperature while the others do not, consistent with decreasing loopgain allowing them to satisfy the thermal stability criterion that D,E,F,G have already met.}
\end{figure}

While the few E and G types that passed the normal resistance cuts had long mean lengths, their yields were very low, and most of their failures displayed an IV pattern unique to these types.
These are the types that had vias down to the titanium to make direct electrical contact.
The D and F types on the other hand had high yield, and very wide biasable regions, without the characteristic instability of the baseline A-type detectors.
The D and F types have no direct electrical contact between palladium-gold and titanium, indicating that the limiting thermal resistance in the A-type is the silicon nitride island thickness, and the electron-phonon relaxation time in our titanium may be shorter than 1-3 microseconds.

\section{Bolometer impedance}

To explore the mechanism of instability seen in the IV curves, we measured the
electrical impedance of the bolometers.
Like Lindeman et al\cite{lindeman2004}, we add a known perturbation to the TES
bias voltage, and measure the response current, though we use a small
amplitude square wave modulation rather than white noise.
The frequency of the modulation was swept from 2.6Hz to 1.6kHz for several bias voltages over the same range used for the IV curves.
Assuming linearity, the amplitude and phase of the bolometer current in the first harmonic provides a measurement of the complex impedance of the bolometer at the input square wave frequency.

We fit the complex impedance data to a model for the admittance $Y$ that includes all three mechanisms for instability - ratio of the TES resistance to the shunt resistor, electrical bandwidth of the readout, and thermal decoupling.
Like the standard model for the ideal TES, this has two time dependent degrees of freedom, the current in the TES and the temperature of the island.

The difference is that the TES temperature ($T_{TES}$) can heat above the temperature of the island ($T_{island}$) through electrical power dissipated in the TES ($P_e$) flowing through the decoupling link ($\gamma G$):

\begin{equation} T_{TES} = T_{island} + \frac{P_e}{\gamma G} \end{equation}

All the bias points for one bolometer are fit simultaneously, with a single value for $\tau_0$ and $\gamma$, and a value per bias voltage for $\mathcal L$ and $R$.

\begin{figure}
\includegraphics[width=3.37in]{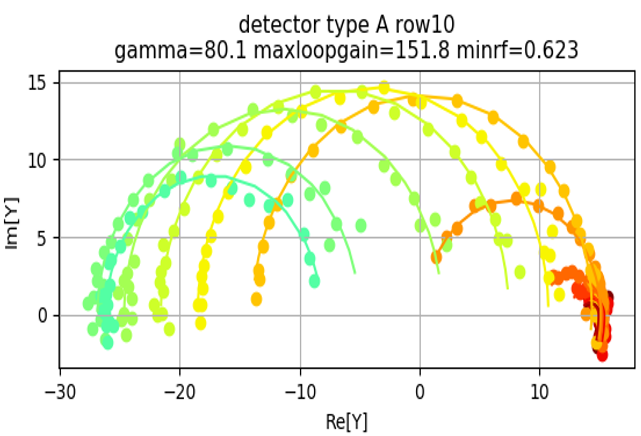}
\includegraphics[width=3.37in]{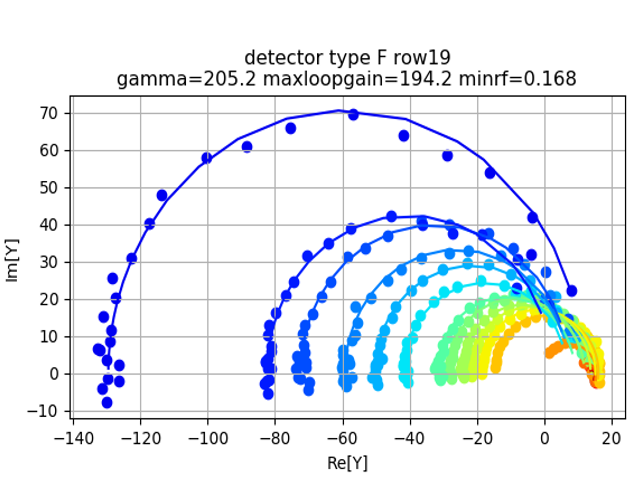}
\caption{\label{fig:admittance} Fitted admittance for representative type A (top) and F (bottom) bolometers.  Points are measured data, one point per frequency, lines are fits to the model.  Color shows the bias voltage.  Data is shown down to the lowest bias voltage where the detector remained stable.}

\end{figure}

\begin{figure}
\includegraphics[width=3.37in]{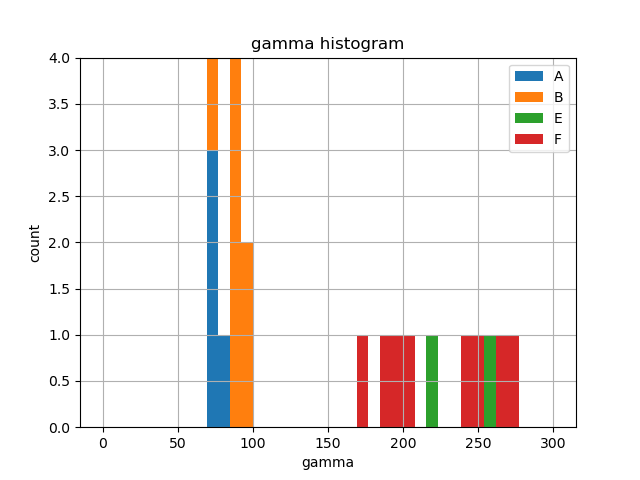}
\caption{\label{fig:gammahist} Histogram of fitted $\gamma$ for the A,B,E and F detectors.} 
\end{figure}

The admittance measurement was performed for detector types A,B,E and F.
Representative plots of the admittance for an A and F type bolometer are shown in Fig. ~\ref{fig:admittance}.
For a detector in the transition, the admittance starts out negative real, then moves in a circle to positive imaginary and then positive real impedance.
The histogram of fitted $\gamma$ is shown in Fig. ~\ref{fig:gammahist}. The A-type measurements are consistent with the expected limiting thermal resistance calculated from Eq. \ref{eqn:limitingThermalConductance}, and the histogram indicates about a factor of three improvement in the internal thermal conductivity between the A and F type bolometers.

\section{Conclusion}
The difficulty in operating our 270GHz detectors is consistent with low internal thermal conductivity in the bolometer, likely from the small cross-sectional area of the silicon nitride island.
Like George et al, despite the very different detector layout and different TES material (manganese doped aluminum), we find that the thermal conductivity could be increased by a factor of 2-3 by bringing the heat capacity metal closer to the TES film.
The stabilized detectors better accommodate variations in optical load and detector properties, in principle allowing a multiplexed column of detectors to be operated at a single bias voltage with higher on sky yield.

In the future we plan to investigate whether the excess high frequency noise observed in Kernasovskiy et al\cite{kernasovskiy2012} can be explained by the internal thermal resistance, and whether the modified detector types reduce the high frequency noise, potentially allowing for reduced sampling rates and higher multiplexing factors.

\section{References}

\bibliography{final_paper_2}

\end{document}